\documentclass[report]{elsart}
\input{psfig.sty}

\usepackage{amssymb}
\input{colordvi.sty}
\newcommand {\pom} {I\!\! P}

\begin{document}
\begin{frontmatter}



\title{Pomeron intercept and slope: the QCD connection}


\author{Konstantin Goulianos}\ead{dino@rockefeller.edu}\ead[url]{http://physics.rockefeller.edu}

\address{The Rockefeller University, 1230 York Avenue, New York, NY 10021, USA}

\begin{abstract}
The ratio of intercept to slope of the Pomeron trajectory is derived in a phenomenological model based on a QCD approach to diffraction.
\end{abstract}

\end{frontmatter}

\tableofcontents
\newpage
\section{Introduction}
Hadronic diffraction has traditionally been treated in the framework of Regge theory\cite{Collins,Barone,Donnachie}. 
In this approach, diffractive processes at high energies are formally described by the exchange of the {\em Pomeron trajectory}, presumed to be formed by a {\em family}
of particles carrying the quantum numbers of the vacuum. Although no particles 
were known to belong to this family, 
the Pomeron trajectory 
was introduced in the 1970s to account for the observations that 
the $K^+p$ cross section was found to be 
increasing with energy at the Serpukov 
70 GeV ($\sqrt{s}=11.5$~GeV for $pp$ collisions) proton synchrotron, and the elastic and total $pp$ cross sections,
which at low energies were falling with increasing energy, started to flatten out and then began 	to rise as collision energies up to $\sqrt s$=60~GeV became available
at the Intersecting Storage Rings (ISR) at CERN. 

In the long history of hadronic diffraction spanning a period of nearly a half century, the intercept $\alpha_{\pom}(0)=1+\epsilon_{\pom}$ of the Pomeron trajectory, $\alpha_{\pom}(t)=\alpha_{\pom}(0)+\alpha_{\pom}'t$, as determined from elastic and total $pp$ and $p \bar p$ cross sections, increased from an initially proposed value of unity to the value of $\alpha_{\pom}(0)\approx 1.08$~\cite{DL}, while the slope parameter $\alpha_{\pom}'$ gradually decreased from $\sim1$~(GeV/c)$^{-2}$ at $\sqrt s\sim 5$~GeV (see~\cite{physrep}) to reach a stable value of $\alpha'_{\pom}\approx 0.25$~(GeV/c)$^{-2}$ at $pp$ and $\bar pp$ collider energies~(see \cite{Donnachie}). In contrast, the Reggeon trajectories formed by the known mesons and resonances have $\alpha_{R}'\approx 1$~(GeV/c)$^{-2}$. To date, no particle or resonance has yet been positively identified to lie on the Pomeron trajectory. The small value of $\alpha_{\pom}'$ relative to $\alpha_{R}'$ remains a theoretical puzzle, whose phenomenological interpretation might contribute to our understanding of the underlying QCD nature of the Pomeron. In this paper, we present a phenomenological model that relates $\textstyle \epsilon_{\pom}$ to $\alpha_{\pom}'$ based on a parton model approach to diffraction.

\section{Regge approach}
In the Regge approach, high energy hadronic cross sections are dominated 
by Pomeron exchange. For $pp$ interactions, the Pomeron exchange contribution to total, elastic, and single diffractive  
cross sections is given by
\begin{equation}
\sigma^{tot}(s)=\beta^2_{\pom pp}(0)
\left(\frac{s}{s_\circ}\right)^{\textstyle\alpha_{\pom}(0)-1}
\Rightarrow
\sigma_\circ{\left(\frac{s}{s_\circ}\right)}^{\textstyle\epsilon_{\pom}},
\label{eq:total}
\end{equation}

\begin{equation}
\frac{d\sigma^{el}(s,t)}{dt}=\frac{\beta^4_{\pom pp}(t)}{16\pi}\;
{\left(\frac{s}{s_\circ}\right)}^{\textstyle 2\left[\textstyle\alpha_{\pom}(t)-1\right]},
\label{eq:elastic}
\end{equation}

\begin{equation}
\frac{d^2\sigma_{sd}(s,\xi,t)}{d\xi dt}=
\underbrace{
\frac{\beta_{\pom pp}^2(t)}{16\pi}\;\xi^{1-2\alpha_{\pom}(t)} }_{\displaystyle f_{\pom/p}(\xi,t)}\;
\underbrace{
\beta_{\pom pp}(0)\,g(t)
\;\left(\frac{s'}{s_\circ}\right)^{\alpha_{\pom}(0)-1}}_{\displaystyle\sigma^{\pom p}\left(s',t\right)}.
\label{eq:diffractive}
\end{equation}
 The differential diffractive cross section, Eq.~(\ref{eq:diffractive}), consists of two terms: the term on the right, 
$\sigma^{\pom p}\left(s',t\right)$, which may be viewed as the $\pom$-$p$ total cross section, and the term on the left, $f_{\pom/p}(\xi,t)$, which  is interpreted as the Pomeron flux emitted by the diffractively scattered proton~\cite{ingeschlein}. 
The parameters appearing in Eq.~(\ref{eq:diffractive}) are defined below:\\
(i) $\alpha_{\pom}(t)=\alpha_{\pom}(0)+\alpha_{\pom}'t=(1+\epsilon_{\pom})+\alpha_{\pom}'t$ is the Pomeron trajectory;\\
(ii) $\beta_{\pom pp}(t)$ is the coupling of 
the Pomeron to the proton usually expressed as 
$\beta_{\pom pp}(t)=
\sigma_\circ \cdot e^{b_\circ t}$, where $\sigma_\circ\equiv\beta_{\pom pp}(0)$ and  
$e^{b_\circ t}$ is a simple exponential expression for the 
form factor of the diffractively escaping proton, $F^2_p(t)=e^{b_\circ\cdot t}$;\\
(iii) $g(t)$ is the triple-Pomeron ($\pom\pom\pom$) coupling;\\
(iv) $s'\equiv M^2$ is the $\pom$-$p$ c.m.s. energy squared, where $M$ is the mass of the diffractively excited proton;\\ 
(v) $\xi\approx M^2/s $ is the fraction of the momentum of the incident proton carried by the Pomeron; and\\
(vi) $s_\circ$ an energy
scale parameter traditionally set to 1~GeV$^2$. 

In analogy with Eq.~(\ref{eq:total}), the Pomeron-proton total cross section is written as
\begin{eqnarray}
{\displaystyle\sigma^{\pom p}\left(s',t\right)=\beta_{\pom pp}(0)\,g(t)
\;\left(\frac{s'}{s_\circ}\right)^{\alpha_{\pom}(0)-1}}= 
{\displaystyle\sigma^{\pom p}_\circ(t)\;
{\left(\frac{s'}{s_\circ}\right)}^{\textstyle\epsilon_{\pom}}}\label{eq:pomtot},\\
{\displaystyle\sigma^{\pom p}_\circ(t)\Rightarrow \displaystyle\sigma^{\pom p}_\circ},
\label{eq:pomtot_not}
\end{eqnarray}
where in Eq.~(\ref{eq:pomtot_not}) $\displaystyle\sigma^{\pom p}_\circ(t)$ was set to a constant, $\displaystyle\sigma^{\pom p}_\circ$, as it has been shown to be independent of $t$~\cite{cool}.

Regge theory was successful in describing elastic, 
diffractive and total hadronic cross sections at energies up to 
$\sqrt s\sim 60$ GeV, with all processes accommodated 
in a simple Pomeron pole approach, as summarized in Ref.~\cite{physrep} (1983).    
Results from an experiment on photon dissociation on hydrogen~\cite{Chapin} (1985) were also well described by this approach. 
However, the early success of Regge theory was precarious. 
The theory was known to asymptotically violate unitarity, 
as the $\sim s^{\epsilon_{\pom}}$ power law increase of total cross sections 
would eventually exceed the
Froissart bound of $\sigma_T<\frac{\pi}{m_\pi^2}\cdot \ln^2s$, which is based on analyticity and unitarity.

The confrontation of Regge theory with unitarity came at much lower energies  
than what would be considered {\em asymptopia} by Froissart bound 
considerations. 
As collision energies climbed upwards in the 1980s to reach $\sqrt s=630$ 
GeV at the CERN S$\bar pp$S collider and $\sqrt s=1800$ GeV at 
the Fermilab Tevatron $\bar pp$ collider, diffraction dissociation
could no longer be described by Eq.~(\ref{eq:diffractive}), signaling a breakdown of factorization.
The first clear experimental evidence for 
a breakdown of factorization in Regge theory was 
reported by the CDF Collaboration in 1994 (see \cite{CDF_sd}, Sev.~VII). In a measurement of the single diffractive cross section in $\bar pp$ collisions 
CDF found a suppression factor 
of $\sim 5$ ($\sim 10$) at $\sqrt s=$546 GeV (1800~GeV) relative to 
predictions based on extrapolations from $\sqrt s\sim$20 GeV. 


\section{Scaling properties and renormalization\label{sec:renorm}}
The breakdown of factorization in Regge theory was traced back to 
the energy dependence of the single diffractive cross section, $\sigma_{sd}^{tot}(s)\sim s^{2\;\displaystyle\epsilon_{\pom}}$, 
which is faster than that of the total cross section, 
$\sigma^{tot}(s)\sim s^{\displaystyle\epsilon_{\pom}}$,
so that as $s$ increased unitarity would be 
violated if factorization held. 
This can be seen more clearly in the $s^{2\;\textstyle\epsilon_{\pom}}$ dependence of $d\sigma_{sd}(M^2,t)/dM^2|_{t=0}$ of the cross section obtained from Eq.~(\ref{eq:diffractive}) by a change of variables from $\xi$ to $M^2$ using $\xi=M^2/s$:
\begin{equation}
\hbox{ Regge:}\;\;\;
d\sigma_{sd}(M^2,t)/dM|_{t=0}^2\sim s^{2\;\textstyle{\epsilon}_{\pom}}/(M^2)^{1+\textstyle\epsilon_{\pom}}.
\label{reggeM2}
\end{equation}

In 1995 it was shown ~\cite{lathuile95,blois95,R} that unitarization could be achieved and the factorization 
breakdown in single diffraction dissociation fully accounted for by 
interpreting the Pomeron flux 
of Eq.~(\ref{eq:diffractive}) as a probability density and {\em renormalizing} it so that  
its integral over $\xi$ and $t$ could not exceed unity:
\begin{eqnarray}
f_{\pom/p}(\xi,t)\Rightarrow N_s^{-1}\cdot f_{\pom/p}(\xi,t)\label{flux}\\
N_s\equiv \int_{\xi(min)}^{\xi(max)}d\xi\int_{t=0}^{-\infty}dt\,f_{\pom/p}(\xi,t)\sim s^{2\epsilon_{\pom}}/\ln s.
\label{eq:renorm}
\end{eqnarray}
\noindent Here, $\xi(min)=M_\circ^2/s$, where $M_\circ^2=1.4$ GeV$^2$ is the effective threshold for diffraction dissociation, and $\xi(max)=0.1$~\cite{R}. 
As the Pomeron flux integral is $\sim s^{2\epsilon_{\pom}}/\ln s$, the $s$-dependence introduced through the 
renormalization factor $N^{-1}_s$ replaces 
the power law factor $s^{2\textstyle{\epsilon}_{\pom}}$ in Eq.~(\ref{reggeM2}) by $\ln s$ ensuring unitarization:

\begin{equation}
\hbox{ Regge/renorm:}\;\;\;
d\sigma_{sd}(M^2,t)/dM|_{t=0}^2\sim \ln s/(M^2)^{1+\textstyle\epsilon_{\pom}}.
\label{renormM2}
\end{equation}

In the QCD inspired parton model approach presented in Sec.~4, this renormalization procedure eliminates overlapping rapidity gaps caused by 
multiple Pomeron emissions while preserving the $(\xi,t)$, or $(M^2,t)$, dependence of the differential cross section.

\begin{center}
\begin{figure}[h]
\vspace*{-5em}
\centerline{\psfig{figure=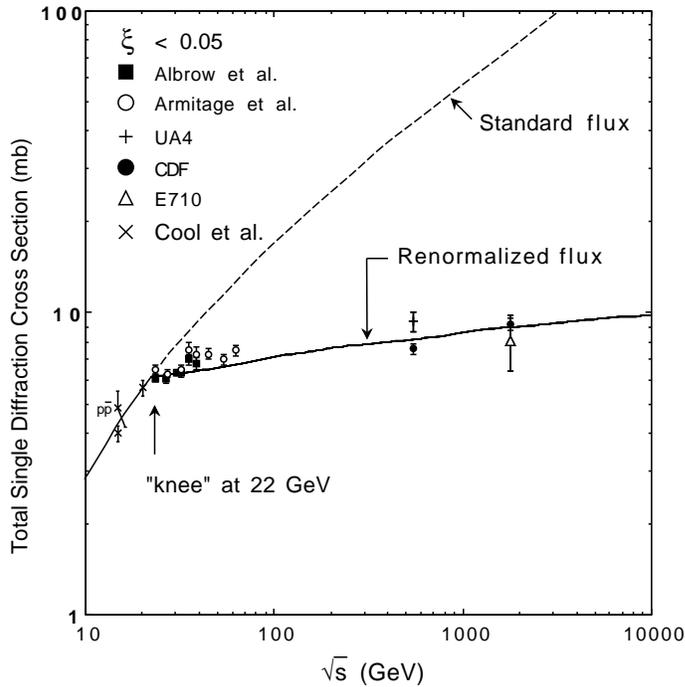,width=0.8\textwidth}}
\vglue -8em    
\caption{\it  Total \protect{$pp/\bar pp$}
single diffraction dissociation cross section data (\,{\em both} $\bar p$ and $p$ sides)   
for \protect{$\xi<0.05$} compared with predictions 
based on the standard and the renormalized Pomeron flux \protect\cite{R}.}
\label{fig:R}
\end{figure}
\end{center}

\begin{center}
\begin{figure}[htp]
\centerline{\psfig{figure=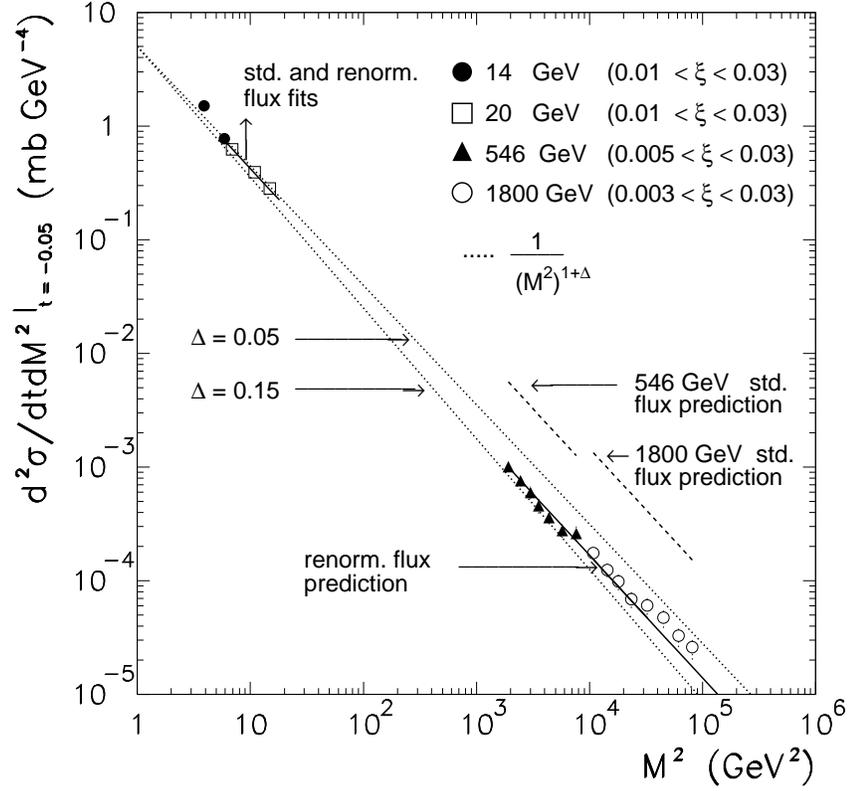,width=0.8\textwidth}}
\vspace*{-1em}
\caption{\it  Cross sections \protect$d^2\sigma_{sd}/dM^2 dt$
for $p+p(\bar p) \rightarrow p(\bar p)+X$ at
$t=-0.05$ GeV$^2$ and $\protect\sqrt s=14$, 20, 546 and 1800 GeV.
Standard (renormalized) flux predictions
are shown as dashed (solid) lines.
At $\protect\sqrt s$=14 and 20 GeV,
the fits using the standard and renormalized fluxes coincide
 \protect\cite{GM}.}
\label{fig:GM}
\end{figure}
\end{center}

In Fig.~\ref{fig:R} (from Ref.~\cite{R}), $\sigma^{tot}_{sd}(s)$ is compared with Regge predictions using the standard or renormalized Pomeron flux. The renormalized flux prediction is in excellent agreement with the data. An important aspect of renormalization is that it leads to a scaling behavior, whereby $d\sigma_{sd}(M^2)/dM^2$ has no power law dependence on $s$. This ``scaling law'' holds for the differential 
soft single diffractive cross section as well, as shown in Fig.~\ref{fig:GM} (from Ref.~\cite{GM}). 

The elastic and total cross sections are not 
affected by the renormalization procedure presented here. Unitarization for the elastic and total cross sections may be  achieved using an eikonal approach, e.g. as reported in Ref.~\cite{CMG} where excellent agreement is obtained between $p^\pm$, $\pi^\pm$, and $K^\pm$ cross section data and the corresponding predictions based on Regge theory and eikonalization. 

The features of the data displayed in Figs.~(\ref{fig:R}) and (\ref{fig:GM}) are obtained below in the parton model approach to diffraction which we use to derive the ratio of $\epsilon$ to $\alpha'$, and thus play a crucial role in validating the model. 

\section{Parton model approach~\label{approach}}
The Regge theory form of the rise of the total $pp/\bar pp$ cross section at high energies, 
$\sigma_{pp/\bar pp}^{tot}(s)=\sigma_\circ\cdot s^{\displaystyle\epsilon}$, which requires a Pomeron trajectory with intercept $\alpha(0)=1+\epsilon$, is precisely the form expected  
in a parton model approach, where cross sections are 
proportional to the number of available ``wee'' (lowest energy) partons. In~\cite{Levin}, the parton model cross section is obtained as $\sigma^{tot}_{pp/\bar pp}=N\times \sigma_\circ$,
where $N$ is the flux of wee partons and $\sigma_\circ$ the cross section of one wee parton interacting with the target proton. The wee partons originate from emissions of single partons cascading down to lower energy partons in tree-like chains. The average spacing in (pseudo)rapidity\footnote{We assume $p_T=1$ GeV, so that $\Delta y'=\Delta \eta'$.} between two successive parton emissions is $\sim 1/\alpha_s$. This spacing governs the wee parton density in the $\eta$-region where particles are produced, defined here as $\Delta\eta'$, which in the case of the total cross section is equal to $\Delta\eta=\ln s$, leading to a total $pp$ cross section of (see \cite{Levin})
\begin{equation}
\sigma_{pp/\bar pp}^{tot}=\sigma_\circ\cdot e^{\displaystyle\epsilon\Delta\eta}.
\label{totDeta}
\end{equation}
This expression is similar in form to the Regge theory Pomeron contribution to the total cross section. 
Since from the optical theorem $\sigma^{tot}_{pp/\bar pp}$ is proportional to the imaginary part of the forward ($t=0$) 
elastic scattering amplitude, the full parton 
model amplitude may be written as 
\begin{equation}
{\rm Im\,f^{el}_{pp/\bar pp}}(t,\Delta\eta)\sim e^{(\displaystyle{\epsilon}+\alpha' t)\Delta \eta},
\label{eq:fPM}
\end{equation}
\noindent where $\alpha' (t)$ is introduced as a simple linear parameterization of the $t$-dependence. 
The parameter $\alpha'$ reflects the transverse size of the cluster of wee partons in a chain, which is governed by the $\Delta\eta$ spacing between successive chains and thereby related to the parameter $\epsilon$.

For the relationship between $\alpha'$ and $\epsilon$ we turn to single diffraction dissociation, which through the coherence requirement isolates the cross section from one wee parton interacting with the proton, since all possible interactions of the remaining wee partons are shielded by the formation of the diffractive rapidity gap.
Based on the amplitude of Eq.~(\ref{eq:fPM}), the single diffractive cross section in the parton model approach takes the form
\begin{equation}
\frac{d^2\sigma_{sd}(s,\Delta\eta,t)}{dt\,d\Delta\eta}=
\frac{1}{N_{gap}(s)}\cdot \underbrace{C_{gap}\cdot F_p^2(t)\left\{e^{\textstyle (\epsilon+\alpha' \,t)\Delta\eta}\right\}^2}_{\textstyle P_{gap}(\Delta\eta,t)}
 \cdot \;\kappa \cdot \left[\sigma_\circ\,e^{\textstyle \epsilon\Delta\eta'}\right],
\label{eq:diffPM}
\end{equation}
where, from right to left:\\
(i) the factor in square brackets represents the cross section due to the wee partons in the $\eta$-region of particle production $\Delta\eta'$;\\
(ii)  $\Delta\eta=\ln s$-$\Delta\eta'$ is the rapidity gap;\\
(iii)  $\kappa$ is a QCD color factor selecting color-singlet $gg$ or $q\bar q$ exchanges to form the rapidity gap;\\ 
(iv) $P_{gap}(\Delta\eta,t)$ is a gap probability factor representing the elastic scattering between the dissociated proton (cluster of dissociation particles) and the surviving proton;\\ 
(v)  $N_{gap}(s)$ is the integral of the gap probability distribution over all phase-space in $t$ and $\Delta\eta$;\\ 
(vi)  $F_p(t)$, in $P_{gap}(\Delta\eta,t)$, is the proton form factor $F_p(t)=e^{\displaystyle b_\circ t}$ defined in the discussion of the parameters that appear in the Pomeron flux in Eq.~(\ref{eq:diffractive}); and\\
(vii) $C_{gap}$ is a normalization constant, whose value is rendered irrelevant by the renormalization division by $N_{gap}(s)$. 

Since $\Delta\eta=-\ln \xi$, the form of Eq.~(\ref{eq:diffPM}) is identical to the Regge form of Eq.~(\ref{eq:diffractive}), identifying $C_{gap}$ and $\kappa\sigma_\circ$ as $\sigma_\circ/16\pi$ and $\sigma_\circ^{\pom p}$, respectively.  The factor $\kappa$ is expressed below in Sec~\ref{sec:ratio}, Eq.~(\ref{eq:kappa}), in terms of the (soft scale) gluon and quark fractions of the proton weighted by the corresponding QCD color factors, ensuring a fully QCD based phenomenological description of the differential single diffraction dissociation cross section on which the derivation of the ratio of slope to intercept rests. 
\vfill 

\section{The ratio \boldmath{$r=\alpha' /\epsilon$\label{sec:ratio}}}
By a change of variables from $\Delta\eta$ to $M^2$ using $\Delta\eta'=\ln M^2$ and $\Delta\eta=\ln s-\ln M^2$, Eq.~(\ref{eq:diffPM}) takes the form
\begin{eqnarray}
\frac{d^2\sigma (s,M^2,t)}{dM^2 dt}=
\left[\frac{\sigma_\circ}{16\pi}\sigma_\circ^{\pom p}\right]
\,\frac{s^{\displaystyle 2\epsilon}}{N(s)}
\;\frac{1}{\left(M^2\right)^{\displaystyle 1+\epsilon}}\;e^{\displaystyle bt}\nonumber\\
\;\;\stackrel{\displaystyle s\rightarrow \infty}{\Rightarrow}\;\;
\left[2\alpha' \,e^{\frac{\displaystyle\epsilon\,b_0}{\alpha' }}
\sigma_\circ^{\pom p}\right]
\frac{\ln s^{\displaystyle 2\epsilon}}{\left(M^2\right)^{\displaystyle 1+\epsilon}}\;e^{\displaystyle bt},
\label{eq:diffM2}
\end{eqnarray}
where $b=b_0+2\alpha'\ln\frac{\displaystyle s}{M^2}$.
Integrating this expression over $M^2$ and $t$ yields the total single diffractive cross section,
\begin{equation}
\sigma_{sd}\stackrel{s\rightarrow \infty}{\rightarrow} 
2\,\sigma_\circ^{\pom p}\;
e^{\frac{\epsilon\,b_0}{2\alpha '}}=\sigma^{\infty}_{sd}\mbox{= constant}.
\label{eq:sigma_not}
\end{equation}

The remarkable property that the total single diffractive cross section becomes constant as $s\rightarrow \infty$ is a direct consequence of the coherence condition required for the recoil proton to escape intact, which selects one out of several available wee partons to provide a color-shield to the exchange and enable the formation of a diffractive rapidity gap. Since diffraction selects the interaction of one of the partons of the outgoing proton, the constant $\sigma\,^{\infty}_{sd}$ is identified as the $\sigma_\circ$ of Eq.~(\ref{totDeta}), which is specific to the dissociating particle, in this case the proton, and therefore equals $\sigma^{pp}_0$. We thus have
\begin{equation}
2\sigma_\circ^{\pom p}\;
e^{\frac{\epsilon\,b_0}{2\alpha '}}=\sigma^{pp}_0,
\end{equation}
which is the sought after relationship between $\epsilon$ and $\alpha\,'$ in terms of constants which can be deduced from QCD parameters through the relationships:
\begin{eqnarray}
\sigma_\circ^{\pom p}=&\beta_{\pom pp}(0)\cdot g(t)=\kappa\sigma_\circ^{pp}\\
\kappa=&\frac{f^{\infty}_g}{N_c^2-1}+\frac{f^{\infty}_q}{N_c}\label{eq:kappa}\\
b_0=&R_p^2/2=1/(2m_\pi^2).
\end{eqnarray} 
Here,  the color factor $\kappa$ is expressed in terms of the $gg$ and $q\bar q$ color factors weighted by the corresponding gluon and sea-quark fractions, and  $R_p$ is the radius of the proton expressed in terms of the pion mass, $m_\pi$. 
The fractions $f^{\infty}_g$ and $f^{\infty}_q$, where the superscript indicates the limit $s\rightarrow\infty$, as in Eq.~(\ref{eq:sigma_not}), are extracted from the CTEQ5L~\cite{CTEQ5L} parameterizations of the corresponding nucleon parton distribution functions $x\cdot f(s)$ at a scale of $Q^2\sim 1$~GeV$^2$, considered the appropriate scale for the soft $pp$ and $p\bar p$ scattering being discussed (see \cite{lathuile04}, Sec.~5.1).
Inserting these parameters in Eq.~(\ref{eq:sigma_not}) yields 
\begin{equation}
\label{eq:r}
r=\frac{\alpha' }{\epsilon}=-[16\,m_\pi^2\ln(2\kappa)]^{-1}.
\end{equation}

The above equation, in which $r$ is expressed in terms of the mass of the pion and the parameter $\kappa$ which depends on QCD color factors and and the gluon and sea-qark fractions of the underlying PDF of the nucleon, represents the sought after ``QCD connection'' between the Pomeron intercept and its slope.  For a numerical estimate of $r$, we use $m_\pi=0.14$~GeV and $\kappa=0.18\pm 0.02$, as obtained for gluon and quark fractions of $f^{\infty}_g=0.75$ and $f^{\infty}_q=0.25$ evaluated from the CTEQ5L nucleon PDF~(see~\cite{lathuile04}, Sec.~5.1). The uncertainty in $\kappa$ is due to an estimated uncertainty of 10~\% in the gluon fraction, with a corresponding uncertainty in the quark fraction as constrained by $f^{\infty}_g+f^{\infty}_q$=1. Using these values yields $r_{pheno}= 3.2\pm0.4$~(GeV/c)$^{-2}$. 

This result is in excellent agreement with the ratio calculated from the values of $\epsilon_{\pom}=0.08$ and $\alpha_{\pom}' =0.25$~(GeV/c)$^{-2}$ for the soft Pomeron trajectory obtained from fits to experimental data of total and elastic $pp$ and $\bar pp$ cross sections for collision energies up to $\sqrt s=$540 GeV, $r_{exp}=0.25\mbox{ (GeV/c)}^{-2}/0.08=3.13$~(GeV/c)$^{-2}$~\cite{DL}.  The smaller value of $r_{exp}$ obtained from a global fit to $p^\pm p$, $\pi^\pm p$, and $K^\pm p$ cross sections, $r_{exp}\mbox{(global fit)}=0.26\mbox{ (GeV/c)}^{-2}/0.104=2.5$~(GeV/c)$^{-2}$~\cite{CMG}, could be attributed to the increase of the intercept due to additional radiation from hard (high $Q^2$) partonic exchanges at higher energies, as for example manifested in the two-Pomeron model of Ref.~\cite{DLhard}. 

\section{Summary}
In a QCD based parton model approach to elastic, diffractive, and total cross sections,
 interactions occur through the emission of partons, which cascade down to wee partons in chains of tree-like configurations. As the spacing between successive emissions is controlled by the strong coupling constant, the total cross section, which is proportional to the number of wee partons produced, assumes a power law behavior similar to that of Regge theory. This partonic description is used to relate the Pomeron intercept of Regge theory to the underlying parton distribution function. The transverse size of the cluster of wee partons in a chain originating from one such emission, which is the source of the slope parameter $\alpha\,'$ of the Pomeron trajectory, depends on the distance in (pseudo) rapidity space between successive emissions and thereby on the parameter $\epsilon$. Exploiting single diffraction, which through the coherence requirement isolates a partonic chain due a single parton emission, 
the ratio of $\alpha'$ to $\epsilon$ is derived in terms of the pion mass, $m_\pi$, and a QCD color factor $\kappa$ appropriately weighted by the gluon and quark fractions of the proton at the soft scale of $Q^2\sim 1$~GeV$^2$, as obtained from the CTEQ5L parameterization~\cite{CTEQ5L} of the nucleon parton distribution function. The derived value of the ratio of $\alpha'/\epsilon$, $r_{pheno}=3.12\pm 0.4$~(GeV/c)$^{-2}$, is in good agreement with the experimental value of $r_{exp}=3.12$~(GeV/c)$^{-2}$.
  



%
\section{Figures}

\end{document}